# Looking at a digital research data archive – Visual interfaces to EASY

Andrea Scharnhorst[1], Olav ten Bosch[2], Peter Doorn[1]

[1]Data Archiving and Networked Services, Royal Netherlands Academy of Arts and Sciences
{andrea.scharnhorst@dans.knaw.nl}
{peter.doorn@dans.knaw.nl}
[2]Drastic Data
{info@drasticdata.nl}

**Abstract.** In this paper we explore visually the structure of the collection of a digital research data archive in terms of metadata for deposited datasets. We look into the distribution of datasets over different scientific fields; the role of main depositors (persons and institutions) in different fields, and main access choices for the deposited datasets. We argue that visual analytics of metadata of collections can be used in multiple ways: to inform the archive about structure and growth of its collection; to foster collections strategies; and to check metadata consistency. We combine visual analytics and visual enhanced browsing introducing a set of web-based, interactive visual interfaces to the archive's collection. We discuss how text based search combined with visual enhanced browsing enhances data access, navigation, and reuse.

## 1 Introduction

In the age of digital libraries, OPAC's (Online Public Access catalogue) determine how the user engages with the collection of a library or an archive [1]. Lists of hits are the preferred form to present retrieval results. Sometimes, the user is also provided with some statistical information – such as overall number of hits, or hits in a certain subject category (presented in a table form). Much more seldom, visual exploration of the databases behind the OPAC is used to give the user feedback on her or his search and to invite her or him to further explore the collection. There is still an obvious gap between a physical encounter with a collection and browsing a collection on-line. Imagine, a library user walking along open stacks or browsing with fingers through card stack drawers of subject catalogs, and compare this with looking at a web-interface. The same holds for the interface to the digital archive EASY at DANS, which offers standard searching and browsing fea-tures[1]. Having said this, there is a large body of literature on new interfaces for digital libraries and collections [2]. Based on this, for some collections, principles as tag clouds[2] [3], network visualizations ("Aquabrowser")[3] [4], or concept browsing[4] (The National Digital Library of the US) have been implemented into search interfaces. Inspired by such experiments we aim to explore the collection of the DANS archive visually with the long-term goal to develop visual browsing tools enhancing search and navigation.

Our research builds on achievements in the information visualization of scientific communication. So-called Science maps [5] display the knowledge space of scientific disciplines, documented in large bibliographic databases such as PubMed[5] or the Web of Knowledge[6]. They also allow overlaying specific activities of institutions, communities, and individuals over bird-eye views of science. In analogy, one can imagine to overlay the resulting hits for a search query of a user on a map of the overall collection. The selection of the attributes (metadata) for such a map, and of the visual language used is complex. It depends essentially of the purpose of the map making [6]. Metadata such as subject categories or classifications are useful for a first orientation in a complex information spaces [7]. So, we build our exploration of EASY around them.

Visualization of data is a high craft, but the last decade has also seen a movement to popularizing and democratizing visualization methods, and introduce interactive visual analytics.[7,8] A large variety of visualization modules and tools are available. One of the author has successfully applied dynamic visualization techniques to a dataset hosted by EASY – the Dutch census data of 1899[9]. In this paper we apply these principles to the whole set of metadata of EASY. Consequently, parallel with the snapshots displayed in this paper we publish the project with the different interactive parts on the web.[10]

---

[1] htttps://easy.dans.knaw.nl/ui/home
[2] An example is the use of a tag cloud in http://www.librarything.com for navigation.
[3] http://zoeken.oba.nl/?q=garden&x=0&y=0
[4] http://strandmaps.nsdl.org/
[5] http://www.ncbi.nlm.nih.gov/pubmed/
[6] http://wokinfo.com/
[7] http://www-958.ibm.com/software/data/cognos/manyeyes/
[8] http://www.gapminder.org/
[9] http://www.drasticdata.nl/ProjectVT/
[10] http://drasticdata.nl/DDHome.php?m=514

## 2    The self-archiving system EASY of DANS

DANS is the largest national research data archive in the Netherlands in the social sciences and humanities. It is a public institution funded in 2005 by the Netherlands Organization for Scientific Research (NWO), and the Royal Netherlands Academy of Arts and Sciences (KNAW). Among its ancestors were economic-historical and sociological archives starting in the 60s already. This long history of a relatively new institution explains some specificity of the metadata. The list of fields or categories used to index datasets is a subset of a disciplinary classification system designed for the Dutch Research Information System (NOD) in the 1980s, and nowadays maintained by another service of DANS - NARCIS, a research information system.

DANS hosts data archives of other scientific institutions based on bilateral agreements. However, it also provides a self-archiving service for individual researcher with a web-interface. Eventually, all archived data sets (deposited by individuals and by institutions) are accessible via EASY. The mission of DANS is to promote sustained access to digital research data. Professional information and documentation (including metadata standards as Dublin Core), storage management and certification processes of so-called trusted repositories are part of its daily practice. At the same time, DANS is a living organization and EASY is a growing database that undergoes changes. In Summer 2012, a new version of EASY was released. In this paper we analyzed metadata extraction at two points in time − later called EASY I set, and EASY II set.

To enable self-archiving, DANS provides detailed instructions on how to deposit data, tailored for social sciences, history and archeology as the main contributing scientific fields and one general instruction[11]. These texts guide the depositor through the EASY interfaces and explain which metadata to provide, which are preferred file formats, what access right options are available, and so on. Users are also asked to choose one or several audiences for the dataset. This information, used to tag the datasets with categories pointing to scientific fields, is meant to stimulate re-use of data and to promote the deposit and sharing of new datasets among certain communities. The self-archiving process is self-guided, but an archivist checks each incoming dataset prior to its publication. This check encompasses the right use of metadata categories as well as a check of the institution of the depositor.

---

[11]  http://www.dans.knaw.nl/en/content/general-instructions-depositing-data

## 3  Data and methods

Two data sets have been used. DANS provided a dump of the EASY I metadata content on November 25, 2011, together with a description of the hierarchical classification of categories used to organize the datasets. Datasets in EASY can be assigned to multiple categories. In the dump these datasets occur multiple times. We found that there were actually 19659 unique datasets in the EASY I set.

EASY II supports the "Open Archives Initiative Protocol for Metadata Harvesting (OAI-PMH)"[12]. We used this to extract the metadata from EASY II on January 20, 2012. As an example, here is the metadata of one such dataset:

```
<aaa:record xmlns="http://www.openarchives.org/OAI/2.0/">
 <aaa:header>
  <aaa:identifier>oai:easy.dans.knaw.nl:easy-dataset:29142</aaa:identifier>
  <aaa:datestamp>2012-01-12T10:27:57Z</aaa:datestamp>
  <aaa:setSpec>D30000:D34000:D34200</aaa:setSpec>
 </aaa:header>
 <aaa:metadata>
  <oai_dc:dc         xsi:schemaLocation="http://www.openarchives.org/OAI/2.0/oai_dc/ http://www.openarchives.org/OAI/2.0/oai_dc.xsd"
xmlns:oai_dc="http://www.openarchives.org/OAI/2.0/oai_dc/"
xmlns:dc="http://purl.org/dc/elements/1.1/">
   <dc:coverage>Brabant, Flanders, Holland</dc:coverage>
   <dc:coverage>1404 - 1482</dc:coverage>
   <dc:creator>Dr R. Stein, Universiteit Leiden, Vakgroep Geschiedenis</dc:creator>
   <dc:date>1996-02-05</dc:date>
   <dc:date>2007-01-31</dc:date>
   <dc:identifier>NHDA: R0104</dc:identifier>
   <dc:identifier>twips.dans.knaw.nl-3570458965826643767-1170150585757</dc:identifier>
   <dc:identifier>urn:nbn:nl:ui:13-86i-k0w</dc:identifier>
   <dc:identifier>easy-dataset:29142</dc:identifier>
   <dc:rights>NO_ACCESS</dc:rights>
   <dc:rights>accept</dc:rights>
   <dc:subject>prosopography</dc:subject>
   <dc:title>Integration from above: the Burgundisation of the Netherlands</dc:title>
   <dc:title>Integratie van bovenaf: de Bourgondisering van de Nederlanden.</dc:title>
  </oai_dc:dc>
 </aaa:metadata>
</aaa:record>
```

The "setSpec" tag refers to a category in the category tree. In this example the dataset is tagged with one category, for which the location in the classification tree is given. In general, a dataset may have multiple setSpecs tags. From the metadata descriptions, we extracted the assignment of datasets to categories, the titles, the creators (could be more than one per dataset), the access rights, the internal EASY identifiers and the persistent identifier.

---

[12] http://www.openarchives.org/OAI/openarchivesprotocol.html

Both metadata sets were cleaned and further processed. The OAI export (EASY II set) also contains metadata from datasets that were deleted from EASY. These were excluded from our analysis. The resulting metadata file contained 21.303 unique datasets.

The EASY categories are hierarchically organized into a tree of max depth 3, which we retrieved from a separate file. Categories can be parents as well as leaf nodes. Leaf nodes may have depth 1, 2 or 3. A category may contain multiple datasets. The EASY I category hierarchy together with the number of datasets directly assigned to each category is shown in Figure 1.

## 4 Results

### 4.1 Some basic statistics

Currently, EASY contains about 20000 datasets. A dataset can contain different files with different file types and of different size. The total number of categories referred to by these datasets is 47. About 1700 unique depositors (creators) can be identified. About 70% of the datasets belong to archeology. About 80% of all datasets are only allocated to one category. Among them are all archeology datasets. Maximal 9 categories are used to tag a dataset.

### 4.2 The category system of EASY

The five main categories currently used in EASY are: *Humanities, Social-sciences, Behavioural Sciences, Social-cultural sciences, Life sciences and Medicine, and Geospatial sciences* (the first level in Figure 1). A user depositing a data set and specifying the relevant categories sees all categories in a dropdown window. That means that the labels of the classes of the category tree are used as controlled vocabulary, but they can in principle be freely assigned. Note that the numbers in Figure 1 do not add up. They describe the actual assignments of datsets to categories. We found datasets that have been tagged with a lower level category in one branch and a higher-level category in an-other branch. Though, we never found that, inside of one branch, categories from different levels are combined.

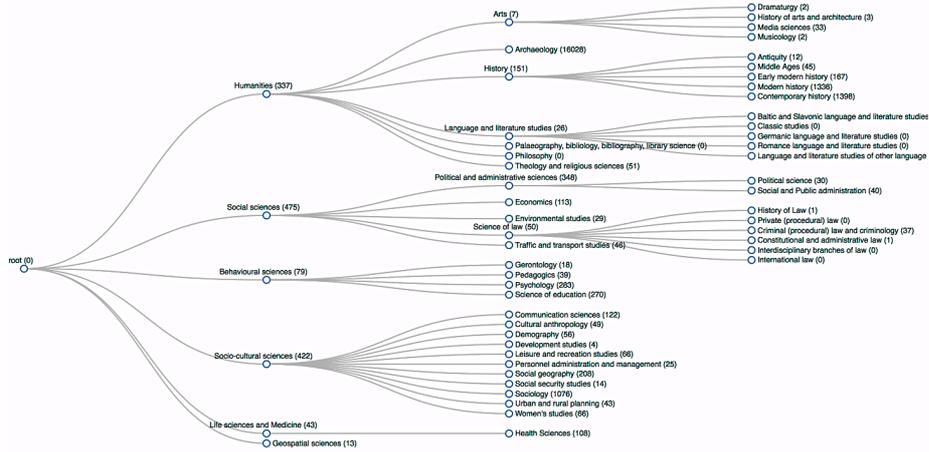

**Figure 1:** Category tree of EASY: The numbers behind the labels show how often this category has been used to tag a dataset (EASY I)

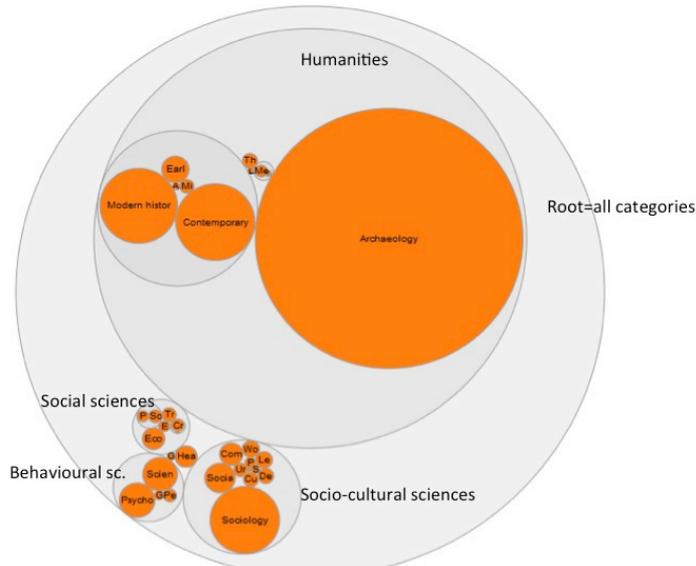

**Figure 2:** Size of categories in EASY (EASY I set)

From a perspective of indexing, it is reasonable that a user might want to first tag a dataset specifically with one category and add another pointing to a wider second audience. For an overview how the tree of categories is actual populated however, one needs to add up the numbers. This has been done in a so-called circle-packing visualization of the occupation (Figure 2). One can see that *Humanities* is the biggest top category basically due to the large amount of datasets in *Archaeology*. *History* is the second largest group inside of humanities. The second largest group on the top level contains datasets tagged as *Social-Cultural Sciences* and/or categories of it, among which *Sociology* is mostly used.

As said above datasets can be tagged with different categories (at different levels). We created an interactive treemap where all datasets with more than one category tag are displayed according to the number of attached categories.[13] This treemap is combined with the visualization of the category tree, on which the size of the category is indicated by a transparent size-scaled circle. (Figure 3) In its interactive counterpart, moving the mouse using over a data set on the treemap, highlights its allocation(s) on the tree. Additionally, the name of the dataset is displayed in a footer below, and clicking on the dataset leads to the original dataset in the archive.

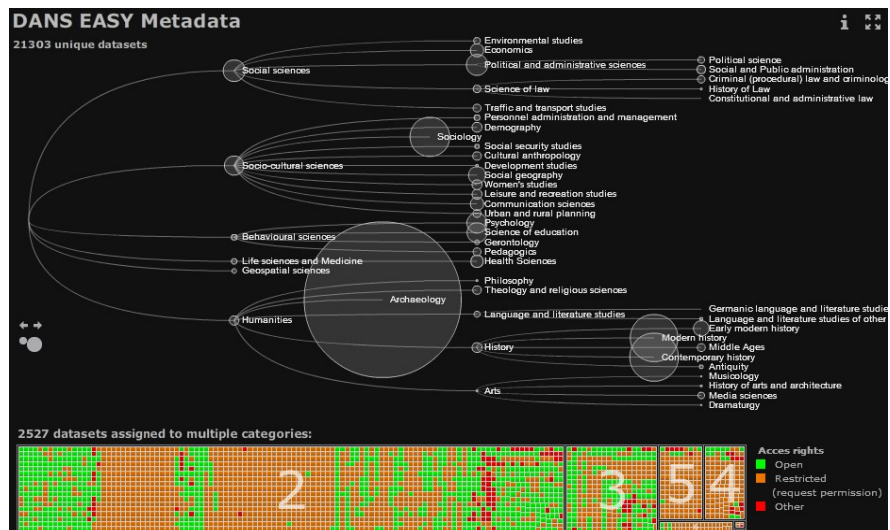

**Figure 3:** A combination of tree and treemap in a visual exploration (EASY II set) – snapshot from http://drasticdata.nl/DDHome.php?m=514

Figure 3 also shows another feature of the dataset – the access rights, which we discuss in more detail in the next section.

---

[13] http://drasticdata.nl/ProjectDANSEasy/indexMultipleAssignments.htm

## 4.3 Access right settings in different disciplines

EASY applies an open access policy and encouraged depositors to make their datasets public. However, specific communities have different norms and practices. Often depositors want to release that data only after a certain time, or deposit data that can be shared on request only. In EASY datasets can be marked in four categories: Open access, Restricted for a specific group, Restricted and Other.

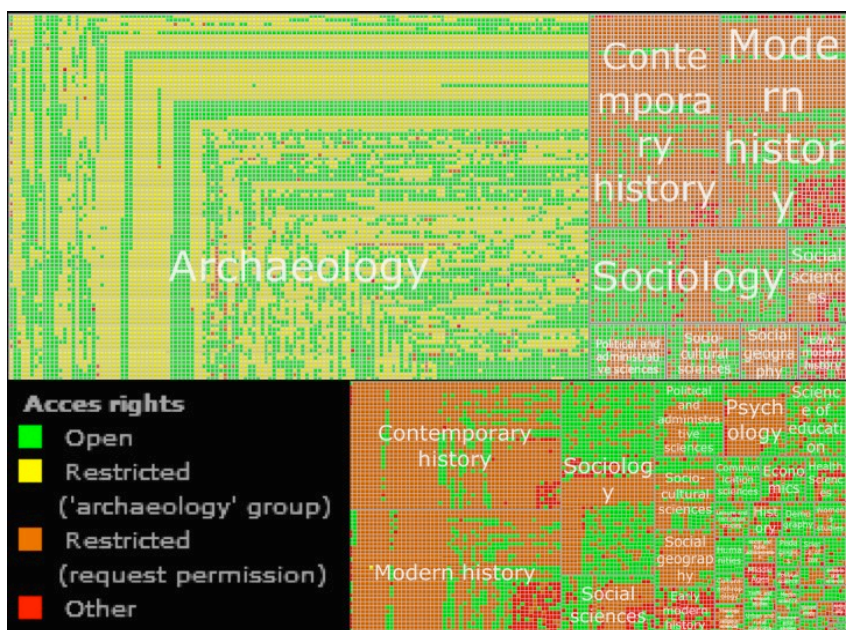

**Figure 4:** Combination of different treemaps of datasets across categories (with all categories above; without Archeology beneath) (EASY II set): snapshots from http://drasticdata.nl/ProjectDANSEasy/indexAllPerCategory.htm

Figure 4 combines snapshots from different treemaps on which datasets are grouped in categories, and color-coded according to access rights. Although the treemap is not nested and so the hierarchy among the categories is washed out, it gives another visual expression of differences in size among fields represented with data in the archive. Be aware that datasets that are assigned to multiple categories are shown multiple times. The treemap (including *Archeology*) has 24.993 items, but only 21.303 of them are unique datasets.

The overall picture reveals that most datasets in the archive oblige the open access policy of DANS. The category "Restricted access for a group" (yellow) has only been applied for datasets in archeology. A treemap without archeology shows clearly in which fields datasets are not immediately open (red).

The snapshots in Figure 4 are taken from different experiments which can be found here: http://drasticdata.nl/DDHome.php?m=514.

### 4.4 Main depositors

Eventually, we analyze and display which depositors contribute to the archive. In the treemap in Figure 5, datasets (access color coded) are grouped according to depositing institutions. We can identify a handful major player. The biggest is the "Veteran Institute".

We also see a larger group of depositors that deposit several datasets. At the left lower corner we find all individual depositors with only one dataset. That the archive gives a home to these many individual datasets is one of the successes of EASY. Figure 5 is a static snapshot.

To allow the user also to search for datasets and depositors we created an interface that combines a treemap (without archeology) on the left with a graphic with all depositors on the right. In this display, a size-coded circle represents each creator. Mousing over a creator (right) shows the datasets deposited left. Mousing over a dataset highlights the depositor. A snapshot of it is shown in Figure 6.

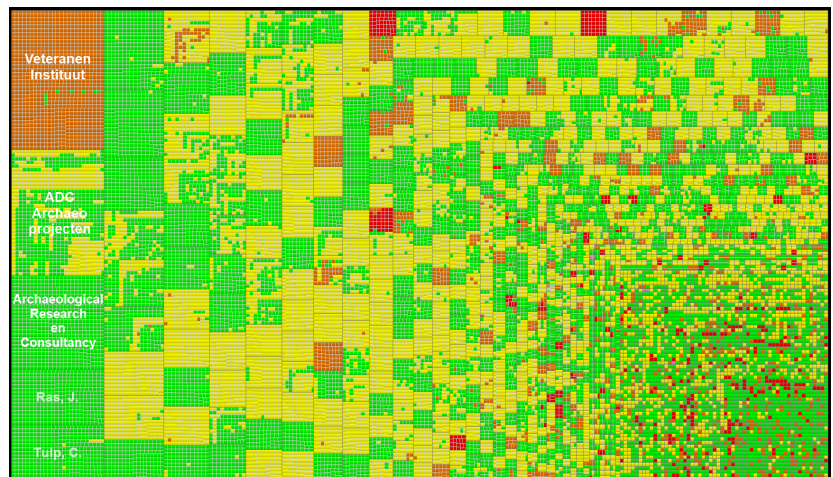

**Figure 5:** Treemap of datasets grouped by depositor (EASY II set)

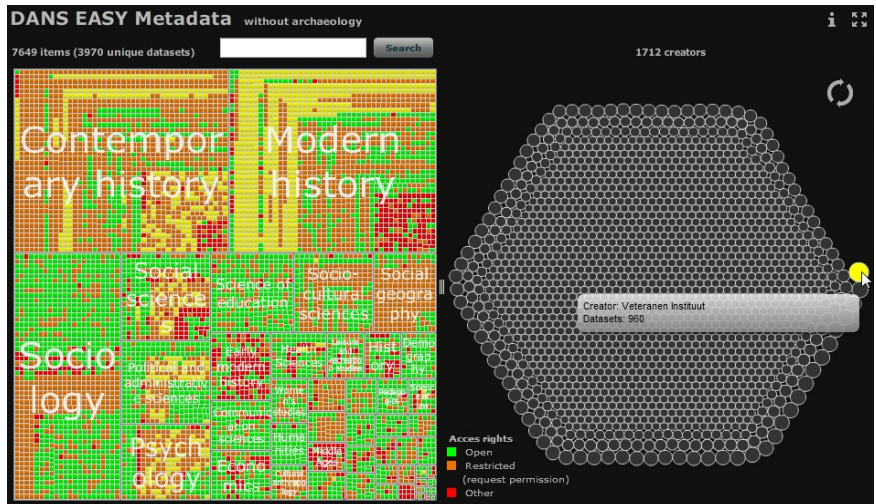

**Figure 6:** Depositors and datasets can be related to each other (EASY II set) – snapshot from http://drasticdata.nl/ProjectDANSEasy/indexAllNoArchaeology.htm

## 5 Conclusions

In this paper we analyzed the metadata of the collection of datasets in EASY, an infrastructure to deposit digitally research data. In particular we analyzed the fields: subject or classification; depositors; and access right. We displayed the structure of the archive's content along these dimensions. Statements about the evolution of the collection in EASY would need analyses of the date field. However, we found that different date fields ("date created", "date available", "date submitted") were mapped into the same Dublin Core field "date". There is another format of export which might allow us to retrieve at least one of the dates, but we left this for further research. This example shows, how an analysis of metadata can help to improve the structure in which those are indexed and made interoperable for harvesting and exchange.

We also discussed the correct interpretation of certain fields, such as the "Driver" value in the setSpec and the interpretation of the access rights with the experts from DANS. The "Driver" value indicates the subset of datasets that have open access. Since the access rights field also describes this, we did a check to see if this was consistent. There were 155 differences. It turned out that they were partly explainable (open datasets still under embargo are not in the Driver setSpec) and partly errors in the metadata. Both examples indicate that analysis of metadata can be used inside the archive to check workflows, optimize interfaces for harvesting, and check for consistencies.

The visual analytics of the depositors is of relevance of the further strategic positioning of the archive. Those actors (together with the downloaders) are important stakeholders. As visible, EASY provides a home of the many individual research data sets that would be lost otherwise.

The visual analysis of the classification system and its use show that the assignment to multiple categories is an essential feature of the metadata of EASY and should not be neglected when developing visual mechanisms.

The presented analysis was also an experiment towards the development of visual facetted and visual enhanced browsing. For this reason we did not only visually analyze the data but published several interactive displays on the web. Each visual representation brings thousands of EASY datasets together into one view. From there, each dataset is only one mouse click away.

We already explained some of the interactive elements. Both the visualization of datasets across categories[14] and across categories and creators[15] entail a search interface. This search function is much more rudimentary than the one used in the interface of EASY. However, it indicates how usual text-based search can be combined with visual exploration.

For this paper we choose metadata dimensions and a visual language that displays the general structure of the collection. Hereby, we only partly explored the richness hidden in the metadata description. To produce maps useful for querying the archive by outside user more research needs to be done into the current user behaviour, for instance based on log-file analysis. Evaluation tests with user groups on pilot interfaces can be a next useful step. Such pilots could be still done with a dump of the metadata. Ultimately, we would like to incorporate such functionalities into the "living" EASY system.

---

[14] http://drasticdata.nl/ProjectDANSEasy/indexAllPerCategory.htm
[15] http://drasticdata.nl/ProjectDANSEasy/indexAllNoArchaeology.htm

**Acknowledgement:** This research has been funded as a KDP (Klein Data Projekt) of DANS. Special thanks goes to Henk Koning (DANS) who provided the EASY I data set, and gave further hints. We are grateful for comments and inspiring discussions with Katy Börner (SLIS, Indiana University, USA) and Linda Reijnhoudt (DANS).